\documentclass[twocolumn,showpacs,preprintnumbers,amsmath,amssymb,aps,floatfix]{revtex4}

\usepackage{graphicx}
\usepackage{dcolumn}
\usepackage{bm}

 \newcommand{\vek}    [1] {\textrm{\textbf{{#1}}}} 
 \newcommand{\Op}			[1]	{\hat{\textrm{#1}}}
 \newcommand{\mi}         {\textrm{i}}   
 
 \newcommand{\ket}    [1] {| #1 \rangle}
 \newcommand{\bra}    [1] {\langle #1 |}
 \newcommand{\bk}     [2] {\langle #1 | #2 \rangle}
 
 \newcommand{\bok}    [3] {\langle #1 | #2 | #3 \rangle}
 \newcommand{\Nuk}		[4]	{\mbox{${}_{#1}^{#2}\text{#3}^{\,#4}$}}
 \newcommand{\Hel}				{\Nuk{}{3}{He}{}}
 \newcommand{\Heli}				{\Nuk{}{}{He}{}}
 \newcommand{\Tmin}				{\Nuk{}{}{T}{-}}
 
 \newcommand{\eV}{\text{eV}}
 
 \newcommand{\meV}{\text{meV}}
 \newcommand{\Ry}{\text{Ry}}
 
\begin{document}

\title{Final-State Spectrum of~ \Hel\ after \boldmath{$\beta^-$} Decay of Tritium Anions $\Tmin$}

\author{Alexander Stark}
\author{Alejandro Saenz}
 \email{alejandro.saenz@physik.hu-berlin.de}
\affiliation{%
AG Moderne Optik, Institut f\"ur Physik, Humboldt-Universit\"at
         zu Berlin, 
         Newtonstr.\,15, D\,--\,12\,489 Berlin, Germany}%

\date{\today}%

\begin{abstract}
The final-state spectrum of $\beta$ decaying tritium anions $\Tmin$ was 
calculated. The wavefunctions describing the initial $\Tmin$ ground state 
and the final $\Hel$ states were obtained by the full configuration-interaction 
method. The transition probability was calculated within the sudden 
approximation. The transition probability into the electronic continuum 
is extracted from the complex-scaled resolvent and is shown to converge for  
very high-energies to an approximate analytical model probability 
distribution. 
\end{abstract}


\maketitle

\section{Introduction}
\label{sec:intro}
The neutrino rest mass is a very important parameter for cosmology,
astrophysics, and the standard model of elementary particles. The 
existence of neutrinos, already postulated by Pauli and put into a 
mathematical framework of $\beta$ decay by Fermi \cite{nu:ferm34} 
long time ago, was verified by Reines and Cowan in 1956 \cite{nu:cowa56}. 
However, despite the high solar neutrinos flux of about billions per 
$\text{m}^{-2}\text{s}^{-1}$ on earth the answer to the question about 
their rest mass is one of the big unknowns in physics. Since neutrino-flavor 
oscillations have been observed in the late nineties at the Super-Kamiokande 
experiment \cite{nu:fuku98} a non-vanishing neutrino rest mass has to be 
expected. Unfortunately, this type of experiments reveals only mass 
differences between neutrino flavors. 

The presently constructed KATRIN (Karlsruhe tritium neutrino-mass) experiment 
with an expected sensitivity of about $0.2\,\eV$ $(90\% \text{ C.L.})$ should 
have the ability to determine the absolute value for one of the flavors or 
at least a new upper limit to it \cite{nu:katr08}. 
This so called next-generation 
tritium $\beta$-decay experiment is only based on kinematic relations and
energy and momentum conservation. Thus KATRIN provides a model-independent 
direct measurement of the antineutrino rest mass $m_{\bar{\nu}_e}$ (more
accurately the mass of the antineutrino in a given mass-flavor mixture mostly 
attributed to the electronic neutrino). In more detail, 
$m_{\bar{\nu}_e}^2$ will be extracted in a fit procedure from the shape of the
$\beta$ spectrum. Besides the precise measurement of the $\beta$-electron
energy spectrum it is crucial for the mass extraction to know how the $\beta$ 
spectrum is modified by the final-state spectrum of the decay product. As in 
the previous most recent tritium neutrino-mass experiments in Mainz and Troitsk 
the $\text{T}_2$ molecule is chosen as tritium source. $\text{T}_2$ comprises 
a compromise between experimental accessibility and theoretical treatability.  
The final-state spectrum of its decay product ${}^{3}\text{HeT}^{+}$ was 
therefore subject of a number of very detailed calculations 
\cite{nu:kolo85,nu:fack85,nu:jezi85,nu:kolo88,nu:froe93,nu:froe96,nu:saen97b,nu:jons98,nu:jons99}, 
finally accumulating in the one covering the whole energy regime  
\cite{nu:saen00}. Recently, the spectrum was further adapted to specific needs 
(isotope distribution and temperature) of the KATRIN experiment 
\cite{nu:doss06}.
  
Although a high purity of the molecular tritium source is expected for KATRIN, 
the produced $\beta$ electrons can interact with other gas molecules and thus 
produce tritium species different from $\text{T}_2$. One of the expected 
processes is the dissociative attachment
\begin{equation}
  \text{e}^{-}+\text{T}_2\,\rightarrow\,\Tmin+\text{T} 
	\label{eq:disatt}
\end{equation}
where $\Tmin$ formation occurs. Despite the relative small cross section
compared to, e.\,g., the one for vibrational excitation of the $\text{T}_2$ 
molecule, this process is very important. The reason is the higher endpoint 
energy of the $\beta$ spectrum for the decay of $\Tmin$ compared to the one 
of $\text{T}_2$. Due to this fact the occurrence of $\Tmin$ ions leads to a 
systematical error and hence to a possible limitation of the sensitivity of 
KATRIN, if it is not properly accounted for \cite{nu:katr04}.

To the authors' knowledge, there exist so far only two theoretical predictions 
for the final-state spectrum following $\beta$ decay of $\Tmin$. However, 
in \cite{nu:frol98} only transition probabilities to 4 final states are 
reported. Furthermore, the results in \cite{nu:frol98} disagree substantially 
from the ones given in an earlier work \cite{nu:hars93} that was, however, 
also limited to 10 final states. The aim of this work was thus to provide 
a complete final-state spectrum for the decay process %
\begin{equation}
  \Tmin\,\rightarrow\,\Hel+\text{e}^{-}+\overline{\nu}_{\text{e}}\,.
	\label{eq:decay}
\end{equation}
and to shed some light on the disagreeing earlier results.

\section{Method and computational details}
\label{sec:method}
The calculation of the non-relativistic eigenstates of the atomic systems 
is performed within the approximation of an infinitely heavy mass of the 
nuclei, i.\,e.\ $\text{T}\approx\Nuk{}{\infty}{H}{}$ and
$\Hel\approx\Nuk{}{\infty}{He}{}$. 
This is justifiable due to the large mass difference of the nucleus and the 
electrons. The calculation of the final-state spectrum can be performed 
analytically for {\it neutral} T atoms and reveals a negligible mass 
dependence. Hence a large mass dependence is also not expected in the case of 
tritium anions. The non-relativistic Hamiltonian for the two-electron system
has the form (atomic units with $m_e=1$, $e=1$, $\hbar=1$ are used throughout, 
if not specified otherwise): %
\begin{equation}
  \Op{H} = - \frac{1}{2} \left(\Delta_1+\Delta_2\right) 
           - Z \left(\frac{1}{r_1}+\frac{1}{r_2}\right) 
           + \frac{1}{\left|\vek{r}_{1}-\vek{r}_{2}\right|}
 \label{eq:HamiltonO}
\end{equation}
where $Z$ is the charge of the nucleus and $\vek{r}_{i}$ the position vector 
of the $i$\,th electron. Due to the fact that the only bound state of $\Tmin$ 
is a singlet state with angular momentum $L=M=0$ \cite{hill77}, only 
symmetric spatial configuration state functions (CSF) are important,
\begin{eqnarray}%
  \ket{\Phi^{(+)}_k}&=&
  \begin{cases}\displaystyle 
    2^{-1/2}\, \left(\ket{\phi_i}\ket{\phi_j}+\ket{\phi_j}\ket{\phi_i}\right)
    & i \neq j\\
    \ket{\phi_i}\ket{\phi_j} & i=j\,.
  \end{cases}
 \label{eq:Sing}
\end{eqnarray}
To determine the eigenstates and corresponding energy eigenvalues a simple 
expansion in Slater-type orbitals (STO) is used, %
\begin{equation}
   \bk{\vek{r}}{\phi_i} = \sqrt{\frac{(2\zeta_i)^{2n+1}}{(2n)!}}\,
                     \exp(-\zeta_i r)r^{n-1}\,Y^m_l(\vartheta,\varphi)\,.
\label{eq:STO}
\end{equation}
The $n,l,m$ are integer parameters with limitations analogously to the ones 
for the hydrogen quantum numbers and the $Y^m_l$ represent the spherical 
harmonics. The $\zeta_i$ are positive real parameters. An appropriate choice 
of these parameters allows for the achievement of an in principle complete 
coverage of the Hilbert space of the one-particle part of 
Hamiltonian \eqref{eq:HamiltonO}. In the full configuration-interaction (CI) 
method the eigenstates are expressed as a linear superposition of all possible 
symmetry-adapted CSFs %
\begin{equation}
  \ket{\Psi_j(\vek{r}_{1},\vek{r}_{2})} 
                              = \sum_{k}{c_{jk}\ket{\Phi^{(+)}_k}} 
\label{eq:CI}
\end{equation}
that can be formed with the aid of the chosen STO basis. The expansion 
coefficients $c_{jk}$ are determined by solving the generalized 
eigenvalue problem obtained from inserting the wavefunction {\it ansatz} 
of Eq.\,\eqref{eq:CI} into the eigenvalue equation of the Hamiltonian 
\eqref{eq:HamiltonO}.

The final-state spectrum of $\Heli$ is calculated within the sudden
approximation \cite{nu:migd69} that is based on the fact that the escaping 
$\beta$ electron has a much higher velocity than the bound electrons. 
In the analysis of tritium neutrino-mass experiments like KATRIN only the 
$\beta$ electrons with an energy near the endpoint of the $\beta$ spectrum 
at $18.6\,\text{keV}$ are used. Their velocity is clearly much larger than 
the average speed of the bound electrons in $\Tmin$. In fact, the validity 
of the sudden approximation has been demonstrated for T$_2$ 
in \cite{nu:froe96,nu:saen97a,nu:saen97b} where the first-order correction 
terms were derived and explicitly calculated. From those results it is 
apparent that also for $\Tmin$ the sudden approximation is expected to 
be valid within the accuracy required for the analysis of an experiment 
like KATRIN. Nevertheless a brief discussion of possible effects beyond the 
sudden approximation on 
the final-state spectrum is given at the end of this work. 

A basis set of 555 STOs yielding 3481 CSFs in the full CI calculation was 
used to obtain the final results shown in this work. This STO basis set contains 
all possible kinds of orbitals (with restrictions on $l$ and $m$ as 
mentioned above) up 
to the angular quantum number $l=7,\, -7 \leq m \leq 7$. For the optimization 
of the parameters $\zeta_i$ a genetic and several other algorithms
\cite{saha01} were tested . However, none of those algorithms lead to completely 
convincing results. Therefore, the parameters were finally optimized by 
hand. The difficulty of the parameter optimization is due to the requirement 
to construct a basis 
set with high coverage of the Hilbert space while avoiding inaccuracies due 
to numerically caused linear dependencies. With the aim to achieve a uniform  
description of the possible $\Heli$ final states it is favorable to obtain 
a homogeneous and a high density of states in the continuum as well as 
a large number 
of bound states. If a large number of CFSs is used, the optimization of the 
individual $\zeta_i$ values becomes less important, since the full CI method 
leads to a sufficient mixing of the Hilbert space covered by the various 
STOs. Therefore, the parameters $\zeta_i$ were chosen to start in an interval  
between 2 and 3 and to decrease in value for increasing $n$ (for a given $l$). 
This procedure avoids numerical problems and allows the construction 
of a huge, but linearly independent basis set. This basis set is used for 
both the ground state of $\Tmin$ and all final states of $\Heli$. The 
chosen basis leads for $\Tmin$ to the ground-state energy 
$E_0^{\rm T^{-}} = -14.3602\,\eV$ that is only $0.8\,\meV$ above the very 
accurate values in \cite{frol07,king97}. In the case of $\Heli$ the 
adopted basis set yields 16 states below the ionization continuum. Out of 
those 16 states 15 are identified as true physical states, while the 16th 
state is a pseudo state that resembles the remaining infinite number of 
Rydberg states as a consequence that a finite basis set is adopted.   

Within the sudden approximation the transition probability for $\Tmin$ decays 
into bound states of $\Heli$ is simply given by the squared overlap 
\begin{equation}
  P_n = \left| \bk{\Psi_{n}^{\rm He}}{\Psi_i^{\rm T^-}} \right|^2 
	\label{eq:Pif}
\end{equation}
of the initial state $\ket{\Psi_i^{\rm T^-}}$, i.\,e. the $\Tmin$ ground
state, and the final state $\ket{\Psi_{n}^{\rm He}}$, i.\,e. the n-th bound 
state of $\Heli$.
\begin{table} 
 \caption{\label{tab:disP} Population probabilities $P_n$ of the 
   $^1 S$ bound states of helium after the $\beta$ decay of a 
   $\Tmin$ anion. Also given are the corresponding energies $E_n$ 
   (in atomic units) obtained in the present work.}
 \begin{ruledtabular}
 \begin{tabular}{c c c c c}
$n$ 	& $E_n$ 												&	$P_n(\%)$							&	$P_n(\%)$								&	$P_n(\%)$ \\ 
		  &	(this work)\footnotemark[1]		& (this work)			& (Ref.~\onlinecite{nu:frol98}) & (Ref.~\onlinecite{nu:hars93}) \\\hline 
$1$		&	$\mathbf{-2.903}4572$				&	$22.98998$			& $22.993764$				& $19.147$ \\
$2$		&	$\mathbf{-2.1459}527$				&	$46.86960$			& $46.867404$				& $21.149$ \\
$3$		&	$\mathbf{-2.0612}659$				&	$0.01320$				& $0.135$						& $0.27$ \\
$4$		&	$\mathbf{-2.03358}41$				&	$0.18363$				& $0.21$						& $0.143$ \\
$5$		&	$\mathbf{-2.02117}49$				&	$0.09220$				& $-$								& $0.07$ \\
$6$		&	$\mathbf{-2.01456}04$				&	$0.05262$				& $-$								& $0.039$ \\
$7$		&	$\mathbf{-2.01062}35$				&	$0.03275$				& $-$								& $0.024$ \\
$8$		&	$\mathbf{-2.00809}09$				&	$0.02175$				& $-$								& $0.016$ \\
$9$		&	$\mathbf{-2.00636}79$				&	$0.01522$				& $-$								& $0.011$ \\
$10$	&	$\mathbf{-2.00514}07$				&	$0.01111$				& $-$								& $0.008$ \\
$11$	&	$\mathbf{-2.00423}55$				&	$0.00848$				& $-$								& $-$ \\
$12$	&	$\mathbf{-2.00355}07$				&	$0.00666$				& $-$								& $-$ \\
$13$	&	$\mathbf{-2.00302}05$				&	$0.00513$				& $-$								& $-$ \\
$14$	&	$\mathbf{-2.002}5951$				&	$0.00469$				& $-$								& $-$ \\
$15$	&	$\mathbf{-2.0022}570$				&	$0.00273$				& $-$								& $-$ \\\hline
$\sum{P_n}$ & 										& $70.30975$			& $70.206168$				& $40.877$
\end{tabular}
 \end{ruledtabular}
 \footnotetext[1]{The bold digits agree with the results in Ref.\,\onlinecite{naka08}.}
\end{table}

\begin{table*} 
 \caption{\label{tab:dsP} Discretized final-state probability distribution
   $P(E_{i})$ for $\Heli$ following the $\beta$ decay of a $\Tmin$ anion. 
   The mean excitation energies $E_{i}$ are given relative to the ground state of $\Hel$.}
 \begin{ruledtabular}
 \begin{tabular}{c c c c c c c c}
$E_{i}(\text{eV})$ 	  & $P(E_{i})(\%)$ 		&	$E_{i}(\text{eV})$		  &	$P(E_{i})(\%)$			& $E_{i}(\text{eV})$ 	  & $P(E_{i})(\%)$ 		& $E_{i}(\text{eV})$ 	  & $P(E_{i})(\%)$ \\\hline
25.084	&	0.36869	&	50.096	&	0.07054	&	75.086	&	0.09926	&	186.05	&	0.00960 \\
26.081	&	0.32908	&	51.097	&	0.07403	&	76.085	&	0.09172	&	191.06	&	0.00873 \\
27.081	&	0.28924	&	52.100	&	0.08005	&	77.085	&	0.08480	&	196.06	&	0.00796 \\
28.081	&	0.25425	&	53.104	&	0.09041	&	78.086	&	0.07863	&	201.06	&	0.00727 \\
29.082	&	0.22462	&	54.112	&	0.10943	&	80.950	&	0.31969	&	206.07	&	0.00666 \\
30.082	&	0.19976	&	55.125	&	0.14982	&	85.967	&	0.23208	&	211.08	&	0.00612 \\
31.083	&	0.17887	&	56.158	&	0.26863	&	90.980	&	0.17471	&	232.06	&	0.03471 \\
32.084	&	0.16126	&	57.305	&	1.44117	&	95.991	&	0.13529	&	272.24	&	0.01998 \\
33.084	&	0.14632	&	57.863	&	16.80345&	101.00	&	0.10714	&	313.03	&	0.01250 \\
34.085	&	0.13358	&	58.911	&	0.15914	&	106.01	&	0.08644	&	352.94	&	0.00832 \\
35.085	&	0.12267	&	59.972	&	0.02281	&	111.01	&	0.07084	&	392.90	&	0.00581 \\
36.085	&	0.11328	&	61.109	&	0.00924	&	116.02	&	0.05883	&	433.04	&	0.00421 \\
37.086	&	0.10517	&	62.095	&	3.16161	&	121.02	&	0.04942	&	472.98	&	0.00314 \\
38.086	&	0.09816	&	63.026	&	0.13662	&	126.03	&	0.04194	&	513.67	&	0.00239 \\
39.087	&	0.09209	&	64.113	&	0.13095	&	131.03	&	0.03590	&	553.22	&	0.00186 \\
40.087	&	0.08684	&	65.091	&	0.09512	&	136.03	&	0.03098	&	591.30	&	0.00148 \\
41.088	&	0.08233	&	66.100	&	0.11217	&	141.03	&	0.02692	&	630.58	&	0.00119 \\
42.088	&	0.07847	&	67.099	&	0.12245	&	146.04	&	0.02355	&	654.79	&	0.00101 \\
43.089	&	0.07523	&	68.105	&	0.13639	&	151.04	&	0.02071	&	664.12	&	0.00089 \\
44.089	&	0.07256	&	69.171	&	0.26130	&	156.04	&	0.01832	&	663.99	&	0.00081 \\
45.089	&	0.07046	&	70.210	&	0.20446	&	161.04	&	0.01629	&	666.03	&	0.00073 \\
46.085	&	0.06892	&	71.109	&	0.10126	&	166.05	&	0.01454	&	688.91	&	0.00063 \\
47.084	&	0.06803	&	72.101	&	0.12307	&	171.05	&	0.01303	&	776.21	&	0.00048 \\
48.090	&	0.06792	&	73.077	&	0.12211	&	176.05	&	0.01173	&	1550.1\footnotemark[1]	&	0.00358\footnotemark[1] \\
49.098	&	0.06870	&	74.083	&	0.10784	&	181.05	&	0.01059	&	$\sum{P(E_i)}$	&	29.65399
\end{tabular}
 \end{ruledtabular}
 \footnotetext[1]{For energies above $904\,\text{eV}$ the model tail in Eq.~\eqref{eq:tail} was used.}
\end{table*}

To calculate the transition-probability density into continuum states the
complex scaling method is used. It is based on the mathematical development by
Aguilar, Balslev, and Combes \cite{csm:agui71,csm:bals71} as well as Simon
\cite{csm:simo72}. The application of this method leads in practice to a 
simple but powerful modification of the Hamiltonian $\Op{H}$ in 
Eq.~\eqref{eq:HamiltonO}, 
\begin{equation}
  \Op{H}(\theta)=\exp(-2\mi \theta)\Op{T}+\exp(-\mi \theta)\Op{V}\, .
	\label{eq:UHU}
\end{equation}
In Eq.~\eqref{eq:UHU} $\Op{T}$ and $\Op{V}$ are the usual kinetic and
potential energy operators of He, respectively. The complex-scaling angle 
$\theta$ can in principle be chosen arbitrarily within $0^\circ \leq \theta \le
45^\circ$. In the limit of an infinite basis all observables calculated with 
the aid of complex scaling should become independent of $\theta$. Since only 
finite basis sets can be applied in practice, only approximate eigenstates 
can be obtained that may depend on $\theta$. The angle $\theta$ can thus be 
understood as a variational parameter that modifies the adopted basis as can 
be seen from the inverse relation between basis-set exponents 
and the scaling angle discussed, e.\,g., in \cite{csm:saen93a}. A
diagonalization of the Hamiltonian~\eqref{eq:UHU} in the basis described 
by the Eqs.~\eqref{eq:Sing} and \eqref{eq:STO} yields the complex-scaled 
energies $E_j (\theta)$ and wavefunctions $\Psi_j(\theta)$ where the 
latter are still defined by Eq.~\eqref{eq:CI}, but with complex 
coefficients $c_{jk}(\theta)$.  

With the aid of the complex-scaled energies and wavefunctions the 
transition-probability density into the electronic continuum can be extracted from the 
complex-scaled resolvent according to \cite{nu:froe93}%
\begin{eqnarray}
\!\!\!\!\!\!\!\!\!\!\!\!   P(E,\theta) &=& \nonumber \\
       & & \!\!\!\!\!\!\!\!\!\!\!\! \!\!\!\!\!\!\!\!\!\!\!
           \frac{1}{\pi}\text{Im}
           \left\{\sum_{k}\frac{
              \bk{\Psi_i^{\rm T^-}(\theta^{*})}
                 {\Psi_{k}^{\rm He}(\theta)}
              \bk{\Psi_{k}^{\rm He}(\theta^{*})}
                 {\Psi_i^{\rm T^-}(\theta)} 
                           }{E_k^{\rm He}(\theta)-E}\right\} . 
	\label{eq:Pint}
\end{eqnarray}
The $\bra{\Psi(\theta^{*})}$ is the biorthonormal eigenstate to 
$\ket{\Psi(\theta)}$. It is obtained from the latter by a transposition 
and complex conjugation of the angular part, while the radial part is only 
transposed but not complex conjugated.  
The sum over $k$ includes all complex-scaled eigenstates and eigenvalues 
calculated by solving the generalized complex symmetric, but non-hermitian 
eigenvalue problem. As discussed above in the limit of exact eigenstates 
the density $P(E,\theta)$ becomes independent of the complex-scaling 
parameter $\theta$. A variation of $\theta$ for approximate eigenstates 
provides the possibility to determine an optimal $\theta_{\text{opt}}$ 
with highest stability. The best approximation of $P(E,\theta)$ is then
obtained according to%
\begin{equation}
  \left.\frac{\partial P(E,\theta)}
             {\partial \theta}\right|_{\theta_{\text{opt}}}
         =\text{min.}\rightarrow\,P(E):=P(E,\theta_{\text{opt}}).
	\label{eq:opTh}
\end{equation}
Furthermore, the $\theta$ dependence of the spectra gives an indication 
for the convergence of the results.

\section{Results}
\label{sec:results}
In Table~\ref{tab:disP} the calculated transition probabilities for 15 
$^1$S bound states of $\Heli$ are listed. The results reveal that almost 
every second $\Tmin$ decay will end in the first excited state of $\Heli$. 
The next probable final state is the $\Heli$ ground state with nearly $23\%$. 
With a summed probability of $0.45\%$ the higher excited $\Heli$ states 
are rarely populated after $\beta$ decay of T$^-$. The sum over all 
calculated bound states yields $70.3\%$. The summation over all calculated 
states (discrete and discretized continuum states) yields the expected value 
of $100.00\%$, since the same basis is used for initial and all final states, 
but indicates the proper numerical implementation. The excellent agreement of 
the energy eigenvalues at the order of $\mu\text{hartree}$ with the very 
accurate data in \cite{naka08} assures on the other hand the high quality 
of the basis set adopted in the present work and its ability to describe many 
states simultaneously with high precision. A closer view on the energies 
shows that the degree of accuracy of the present results follows the expected 
trends. First, the accuracy increases with $n$, since the importance of 
correlation decreases, if the state becomes more asymmetric and the two 
electrons have smaller spatial overlap. For even higher values of $n$ the
states become increasingly diffuse and thus it is very difficult to describe 
them properly without running into numerically caused linear dependencies. 

A comparison to the final-state probabilities reported by \citet{nu:frol98} and 
\citet{nu:hars93} is also given in Table~\ref{tab:disP}. Especially for 
the highly populated ground and first excited states the results of  
this work confirm the expectedly very accurate results of 
\citet{nu:frol98} that were obtained with explicitly correlated basis 
functions. The agreement for the third excited state ($n=4$) is, however, less  
good, and for the second excited state $(n=3)$ there is even an order of 
magnitude difference. All attempts to improve the basis set for this state 
failed to yield a better agreement. This could be an indication for a 
typographical error (a missing zero after the decimal point) in \cite{nu:frol98}. 

The comparison with  
the results in \cite{nu:hars93} that were obtained with a relativistic MCDF 
(multi-configuration Dirac-Fock) method shows on the other hand pronounced 
differences. The deviation is most remarkably for the first excited state 
that according to the present work and \cite{nu:frol98} should be populated 
with about 47\,\% probability and thus should clearly dominate the final-state 
distribution. However, in the MCDF results in \cite{nu:hars93} its 
probability is found to be about 21\,\%. For the other states, except $n=3$, 
the results in \cite{nu:hars93} are always smaller than the present ones. 
The deviation increases rather uniformly from about 17 to 28\,\% for 
$n$ varying between 1 and 10. Since relativistic effects are expected to 
be small for light nuclei like $\Tmin$ and $\Hel$, it appears very likely 
that the main reason for the difference of the results in \cite{nu:hars93} 
to the present ones (as well as the ones in \cite{nu:frol98}) is due to 
the small number of configurations used in the MCDF method compared with 
the present full CI method. Unfortunately, no details (like energies) of the 
MCDF calculation in \cite{nu:hars93} are available to further clarify 
this issue, but any realistic estimate of the size of relativistic effects 
excludes their responsibility for the large discrepancy between the 
results in \cite{nu:hars93} compared to the non-relativistic calculations  
of this work or the one in \cite{nu:frol98}.  

\begin{figure}%
\includegraphics[width=\columnwidth]{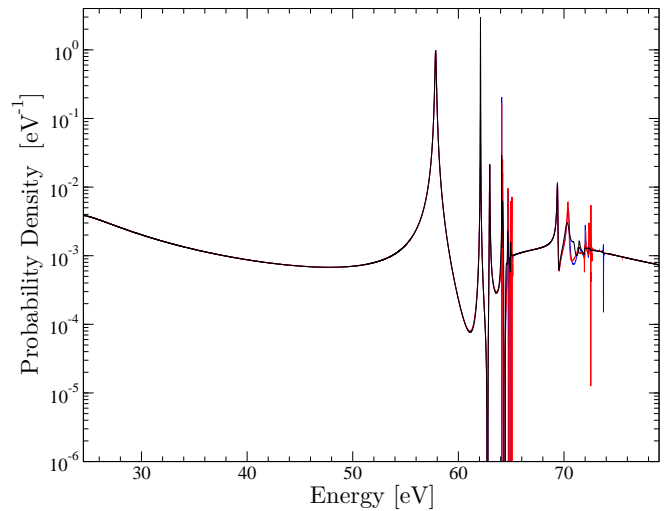}
\caption{\label{fig:spectr}(Color online) Final-state continuum probability 
density of $\Heli$ after $\beta$ decay of $\Tmin$ for $\theta=24^\circ$(red), 
$30^\circ$(blue), and $36^\circ$(black). 
(The energy scale is given relative to the $\Heli$ ground state.)} 
\end{figure}
The calculated transition-probability density into the electronic continuum 
of $\Hel$ is presented for three different complex-scaling angles
($\theta=24^\circ, 30^\circ,$ and $36^\circ$) in Fig.~\ref{fig:spectr}. 
The overall spectrum is practically independent of $\theta$. This indicates 
the high quality of the adopted basis set also for describing the electronic 
continuum. As is usually the case, (higher lying) resonances are most
sensitive to the choice of $\theta$. This is due to the fact that it is 
difficult to find a single value of $\theta$ that is equally appropriate for
describing a certain resonance and the underlying background continuum.  
 
The continuum probability density is dominated by a peak corresponding to 
the first doubly excited singlet state 2s$^2$. About $19\%$ of the $\Tmin$ 
decays ends up in the energy interval between $54.5\,\eV$ and $60\,\eV$. 
Above the $65.4\,\eV$ threshold the higher-lying doubly excited states 2s$n$s  
and in the regime up to $79\,\eV$ (with diminishing importance) the 3s$n$s  
peaks can be identified. The complex-scaling method provides the probability 
density $P(E)$ at any value of $E$ and thus as 
a continuous function. In view of the sharp resonant structures and in 
accordance with the experimental needs, the final-state distribution 
is given in a discretized form as in \cite{nu:saen00}. For this purpose, the 
probability distribution $P(E)$ has been divided into small bins covering an 
energy range of $1.0$\,eV (up to a transition energy of $78.59\,\text{eV}$), 
$5.0$\,eV (from $78.59$ to $214\,\text{eV}$), and $40.0\,\text{eV}$ 
(from $214$ to $904\,\text{eV}$). For each bin the average excitation energy 
$E_i$ and the integrated transition probability $P(E_i)$ were calculated and 
are given in Table~\ref{tab:dsP}.

For the high-energy continuum states (above $904\,\text{eV}$) an approximate 
model tail is introduced, similar to the case of T$_2$ \cite{nu:saen97b,nu:saen00}. 
However, the situation is more complicated for T$^-$. In T$_2$ $\beta$ decay 
the high-energy tail was derived based on the idea that for sufficiently 
large energies of the escaping (formerly bound) electron the effective 
potential of the remaining $^3$HeT$^{2+}$ ion can be well approximated 
by a point charge $Z=2$. In fact, the remaining electron and tritium nucleus 
may be viewed as pure spectators and thus the transition probability should 
approach for high energies the one obtained for a $\beta$-decaying tritium 
{\it atom} for which an analytical result is known. Due to the existence 
of two equivalent electrons, the atomic result is simply multiplied by a 
factor of two \cite{nu:saen97b}. 

While for T$_2$ a hydrogenic wavefunction 
is a reasonable first-order approximation for the initial state, this is 
not the case for T$^-$. In fact, within independent-particle models T$^-$ 
is unstable. As a consequence, the fast electron in the final state may 
be well represented by a Coulomb wavefunction for a point charge $Z=1$ 
(formed by the remaining He$^+$ ion), but the modeling of the initial 
state is less obvious within an independent particle model. This is also 
evident from the alternative point of view that a description of the 
remaining T nucleus and bound electron as a spectator would correspond for 
the active electron to an initial state with $Z=0$ and thus no bound state. 
In order to obtain an atomic-like high-energy tail the initial state is 
thus modeled as a hydrogen-like state with variable exponent. This exponent 
is then obtained by fitting the model spectrum to the ab initio spectrum 
of the full two-electron calculation in the energy range between 500\,eV 
and 10,000\,eV.

The initial $\Tmin$ ground-state wavefunction (omitting the spin 
part for better readability) is then approximated as %
\begin{equation}
  \ket{\tilde{\Psi}^{\rm T^{-}}_i} = \ket{1s^{Z_e}1s^{Z_e}} 
\end{equation}
where $\bk{\vek{r}}{1s^{Z_e}}$ is an STO with $(n,l,m)=(1,0,0)$, i.\,e.\ an atomic 
hydrogen 1s orbital with effective charge $Z_e$. In the spirit of the 
sudden approximation the spectator electron remains in its orbital 
and the final state of the $\Heli^{+}$ ion is modeled as %
\begin{equation}
  \ket{\tilde{\Psi}^{\rm He}(E)} = 
             2^{-\frac{1}{2}}\left(
        \ket{1s^{Z_e}\phi_c(E)}+\ket{\phi_c(E)1s^{Z_e}}\right) 
\end{equation}
where $\ket{\phi_c(E)}$ is the Coulombic continuum wavefunction for energy $E$ 
and charge $Z=+1$.  
Using these model wavefunctions the analytic expression %
\begin{eqnarray}
\!\!\!\!\!\!\!\!\!\!\!\!   \tilde{P} (Z_e,E) &=& \nonumber \\
       & & \!\!\!\!\!\!\!\!\!\!\!\! \!\!\!\!\!\!\!\!\!\!\!
		     2\left(
   \frac{8(1-Z_e)Z_e^{3/2} e^{-2\frac{\arctan({\kappa/}{Z_e})}{\kappa}}}
        {\sqrt{1-e^{-\frac{2\pi}{\kappa}}}\left(\kappa^2+Z_e^2\right)^{2}}
                      \right)^2 \frac{dE}{\Ry} 
	\label{eq:tail}
\end{eqnarray}
with $\kappa=\sqrt{{(E+2)}/\Ry}$ and $1\Ry=13.60585972\,\eV$ is obtained for 
the probability density, i.\,e.\ for the model tail for the 
high-energy continuum states. A fit to the ab initio spectrum yielded 
$Z_e=1.3074$. As is evident from \eqref{eq:tail} the probability density decays 
exponentially for high energies which is the most essential property. (In fact, 
it has been verified that all subsequent conclusions are unchanged, if a 
different model tail is used in which the initial-state charge is fixed 
to the one of the tritium nucleus ($Z=1$) and the effective charge of the 
final Coulomb wave is used as fit parameter.) 

\begin{figure}
\includegraphics[width=\columnwidth]{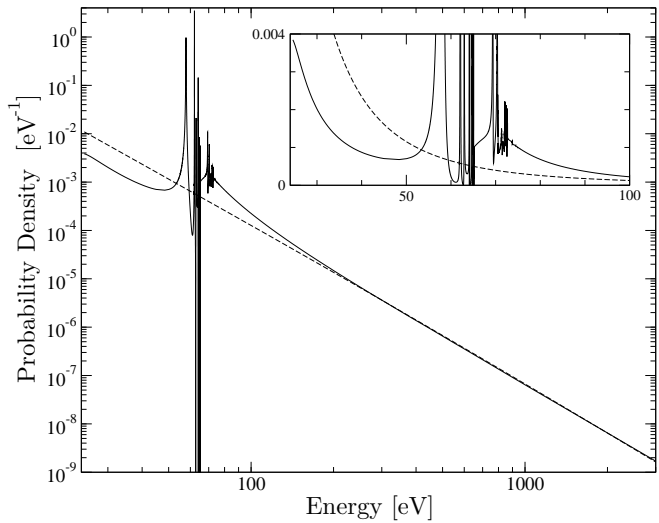}
\caption{\label{fig:tail} Final-state continuum probability 
density of $\Heli$ after $\beta$ decay of $\Tmin$ (solid) and a model tail with $Z_e=1.3074$
(dashed). The inset shows the probability density on a linear scale.}
\end{figure}

Fig.~\ref{fig:tail} compares the calculated 
final-state probability density with the analytical model tail and confirms the 
applicability of the latter for high energies, in fact already starting from about 
200 to 250\,eV, similarly as for T$_2$. The integrated probability density 
(including the model tail for energies above $904\,\text{eV}$) yields a total probability 
of $29.65\%$ for ionization of He following $\beta$ decay of $\Tmin$. 

The mean excitation energy $\overline{E}$ relative 
to the electronic ground-state energy of $\Tmin$ can be obtained from %
\begin{eqnarray}
\!\!\!\!\!\!\!\!\!\!\!\!\!\!\!\!\!\!    \overline{E} & = & \sum_nE_nP_n +\nonumber \\
       & & 
            \int_{24.59\eV}^{904\eV}\,EP(E)\,dE +
            \int_{904\eV}^{\infty}\,E\tilde{P}(E)\,dE .
	\label{eq:meanEsp}
\end{eqnarray}
Insertion of the final-state probability distribution calculated in this work 
in \eqref{eq:meanEsp} yields  
$\overline{E}=27.487\eV$ for the mean excitation energy of the decay product.
This result may be compared to the one obtained by the alternative 
relation
\begin{eqnarray}
  \overline{E} & = & \bok{\Psi_i^{\rm T^-}}{\Op{H}(\Heli)}{\Psi_i^{\rm T^-}} \nonumber \\
               & = & E_0(\Tmin) - 2\,
                     \bok{\Psi_i^{\rm T^-}}{\frac{1}{r}}{\Psi_i^{\rm T^-}}.
	\label{eq:meanE}
\end{eqnarray}
With the expectation value of $\left<r^{-1}\right>$ and the ground state 
energy $E_0 (\Tmin)$ reported by Frolov in \cite{nu:frol98} the mean excitation 
energy calculated with \eqref{eq:meanE} is $\overline{E}=27.469\,\text{eV}$. 
Frolov also reported the expectation values for the case of a finite mass of the 
tritium anion nuclei. Again using \eqref{eq:meanE} for this case one obtains 
$\overline{E'}=27.479\,\text{eV}$. This comparison confirms the 
quality of the final-state distribution obtained in this work and validates furthermore 
the use of the approximation of an infinitely heavy nucleus approximation.

Frolov noted in \cite{nu:frol98} that his calculated bound-state probability 
appears to imply a continuum contribution of about 30\,\%. Since this value is 
about 10 time larger than the continuum probability known for neutral T atoms, 
he speculated that in fact a large number of decays (about 15 to 20\,\%) 
may end up in triplet states of helium, leaving a much smaller fraction in 
the singlet continuum. The reasoning in \cite{nu:frol98} is that the more 
diffuse ground state 
of T$^-$ is not sufficient to explain such a large continuum contribution, 
as follows from a comparison to results obtained for Rydberg states of neutral 
T atoms. On the other hand, triplet states may be populated by the (virtual) 
interaction with the $\beta$ electron omitted in the sudden approximation. 
This argumentation is, however, erroneous. As has been discussed in detail 
in \cite{nu:saen97a}, the sum rule for the sudden approximation gives 
always unity, independently on higher-order corrections to it. This is 
also (as discussed above) fulfilled by the present calculation which 
indeed confirms the about 30\,\% continuum probability indirectly found 
but rejected in \cite{nu:frol98}. 

Finally, the exchange interaction 
is expected to be much smaller than the direct one (in \cite{nu:will83}  
it was found for atomic tritium to be by a factor $\eta^2$ smaller), 
but already the direct term (first term beyond the sudden approximation) 
is by a factor $\eta^2$ smaller than the sudden approximation. Close 
to the end point of tritium $\beta$ decay one finds for the Sommerfeld 
parameter $\eta\approx -0.0271$ 
and for T$_2$ it was explicitly shown that the first-order correction 
to the sudden approximation is itself of the order of $\eta^2$ and thus 
of the order of 0.01\,\% \cite{nu:saen97b}, in accordance with corresponding 
system-independent sum rules given in \cite{nu:saen97a}.

\section{Conclusion}
\label{sec:conclusion}
In this work the complete final-state probability distribution of He following
the nuclear $\beta^-$ decay of tritium anions has been calculated. For the
small number of bound states considered previously in \cite{nu:frol98} the 
agreement is very good for the dominant ground and first excited states, but 
especially for the (very weakly populated) $n=3$ state a deviation by about 
an order of magnitude was found. It is preliminarily attributed to a possible 
typographical error. The agreement with an earlier relativistic
multi-configuration Dirac-Fock calculation \cite{nu:hars93} is on the other 
hand very poor. Since such a large size of relativistic effects is not 
expected, especially not for the light nuclei involved, this deviation 
is attributed to a possibly too small basis set used in \cite{nu:hars93}. 
Nevertheless, the present study may stimulate further theoretical work 
to clarify the discrepancies to \cite{nu:frol98} and of both works 
to the relativistic ones in \cite{nu:hars93}.

In order to further test the accuracy of the present calculation, the
bound-state energies were compared to very accurate literature data and 
were found to agree very accurately with them. Furthermore, the mean 
excitation energy obtained from the complete final-state spectrum was 
compared to the value predicted on the basis of closure. Again, very good 
agreement was found. Therefore, the results of this work should be reliable 
and of direct importance for the tritium neutrino-mass experiment KATRIN 
that is presently under construction. In order to allow the use in the 
experimental analysis and for predicting how much a possible T$^-$ admixture 
to the T$_2$ source spoils the extracted neutrino mass, the continuum 
transition probability is given in binned form, but also available numerically 
on request. Finally, it may be noted that a controlled admixture of T$^-$ to 
the tritium source may in fact be used for the analysis of the experimental 
sensitivity to the atomic and molecular final-state spectrum.    

\begin{acknowledgments}
The authors acknowledge financial support from the {\it Stifterverband 
f\"ur die Deutsche Wissenschaft} and the {\it Fonds der Chemischen Industrie}. 
\end{acknowledgments}


\end{document}